\documentstyle[sprocl]{article}

\def\vep{\varepsilon}

\def\be{\begin{equation}}
\def\ee{\end{equation}}
\def\bea{\begin{eqnarray}}
\def\eea{\end{eqnarray}}

\newcommand \bean {\begin{eqnarray*}}
\newcommand \eean {\end{eqnarray*}}
\newcommand \bary {\begin{array}}
\newcommand \eary {\end{array}}
\newcommand \lan {\langle}
\newcommand \ran {\rangle}
\newcommand \re {\rm Re}
\newcommand \im {\rm Im}
\newcommand{\bi}{\bibitem}
\newcommand{\bit}{\begin{itemize}}
\newcommand{\eit}{\end{itemize}}

\begin{document}

\title{CP Violation}

\author{Lincoln Wolfenstein}

\address{Department of Physics,
Carnegie Mellon University,
Pittsburgh,\\ PA 15213, USA\\
E-mail: lincoln@cmuhep2.phys.cmu.edu} 

\maketitle\abstracts{Three possibilities for the origin of $CP$ violation are discussed: (1) the Standard Model in which all $CP$ violation is due to one parameter in the CKM matrix, (2) the superweak model in which all $CP$ violation is due to new physics and (3) the Standard Model plus new physics.  A major goal of $B$ physics is to distinguish these possibilities.  $CP$ violation implies time reversal violation (TRV) but  direct evidence for TRV is difficult to obtain. }

\section{Introduction}
The symmetries $P$, $C$, and $T$ have played a large role in the
physics of the past 50 years, where $P$ is left-right symmetry, $C$ is
particle-antiparticle symmetry and $T$ for time reversal.  Originally,
these symmetries were not postulated but discovered theoretically as
symmetries of the Hamiltonian of QED.  In nuclear and particle physics
one tried to guess at the Hamiltonian or Lagrangian density without
any classical analogue.  The first of these was the Fermi theory of
the weak interaction governing nuclear beta decays:
\be
\label{fermiint}
{\cal H}_W = G_F\,\left( {\bar p}\,\gamma_{\mu}\,n\;{\bar
e}\,\gamma^{\mu}\,\nu \right)\;\delta(r).
\ee
Indeed this may be said to mark the beginning of particle physics.

As a result of many experiments on nuclear beta decays it appeared
that additional terms would have to be added to Eq.~(\ref{fermiint})
to account for spin-flip (Gamow-Teller) transitions.  The major
advance, however, followed from the observation in 1956 by Lee and
Yang that no experiment had been sensitive to whether parity was
conserved in the weak interaction.  This led to a number of
experiments that showed that $P$ and also $C$ were very much
violated.  This could be accounted for by replacing the fields in
Eq.~(\ref{fermiint}) by left-handed chiral fields \cite{FG1958}, or,
equivalently, replacing $\gamma_{\mu}$ by $\gamma_{\mu}(1-\gamma_5)$.
This became the standard $(V-A)$ theory.

Even before Lee and Yang it had been shown that it was easy to
construct Hamiltonians that violated $C$ or $P$ or $T$ but that in
fact for every relativistic local quantum field theory one could
always define the symmetry $CPT$.  Thus the $(V-A)$ theory violated
$C$ and $P$ but still had the symmetries $CP$ and $T$.  Thus it came
as a great surprise when it was discovered in 1964 that there was a
small violation of $CP$ symmetry in $K^0$ decays.

The $K^0$ is characterized by an additive quantum number $S$ conserved
in strong and electromagnetic interactions but violated by the weak
interaction.  Since $S$ is not a good quantum number $K^0$ with $S=1$
mixes with $\bar K^0$ with $S=-1$, so one expected the eigenstates to
be CP eigenstates
\bean
|K_1\ran &=& \frac1{\sqrt{2}}\left( |K^0\ran + |{\bar K^0}\ran
\right), \\
|K_2\ran &=& \frac1{\sqrt{2}}\left( |K^0\ran - |{\bar K^0}\ran
\right), \\
|{\bar K^0} \ran &\equiv& CP\; |K^0 \ran.
\eean
The observed eigenstates were $K_S$ (lifetime $\tau_S=0.9 \times
10^{-10}$ sec) and $K_L$ ($\tau_L=5.2 \times 10^{-8}$ sec) with
$m_L-m_S=0.48\,\Gamma_S \sim 10^{-5} {\rm eV}$.  The primary decays
were
\bean
K_S &\to& \pi^+\,\pi^-\; {\rm and}\; \pi^0\,\pi^0, \\
K_L &\to& 3 \pi,
\eean
consistent with the CP assignment $K_S=K_1$ and $K_L=K_2$.  The
discovery in 1964 was that $K_L$ also decayed into $\pi^+\,\pi^-$ with
a small branching ratio.

There appeared then two alternatives:
\begin{enumerate}
\item Modify the $\Delta S=1$ interaction in some small way from the
standard $(V-A)$ form.
\item Assume there exists a much weaker $\Delta S=2$ interaction that
violates $CP$.  This would be described by an effective Hamiltonian
(in terms of quarks)
\be
{\cal H}_{sw} = G_{sw}\;{\bar s}\,{\cal O}\,d\;{\bar s}\,{\cal O}\,d +
h.c.,
\ee
where ${\cal O}$ is some Dirac operator.  It is sufficient that
\be
\label{Gsw}
G_{sw} \sim 10^{-10}\;{\rm to}\;10^{-11} G_F.
\ee
\end{enumerate}

The superweak idea is that $CP$ violation is confined to $K^0-{\bar
K^0}$ mixing, which is a $\Delta S=2$ process, second order for the
standard theory.  In the $K_1-K_2$ representation
\bean
M-i\,\frac{\Gamma}{2} = 
\left( \bary{cc}
M_1 & i\,m^{\prime}/2 \\ -i\,m^{\prime}/2 & M_2
\eary \right)
-\frac{i}{2}
\left( \bary{cc}
\Gamma_1 & \\ & \Gamma_2
\eary \right),
\eean
where $m^{\prime}$ is the superweak term and the $i$ is required by
$CPT$ invariance.  Then
\bea
&&\bary{ccc}
|K_S \ran = \left( |K_1 \ran + {\tilde \vep} |K_1 \ran \right)
/ \left( 1+|{\tilde \vep}|^2 \right), \\
|K_L \ran = \left( |K_2 \ran + {\tilde \vep} |K_2 \ran \right)
/ \left( 1+|{\tilde \vep}|^2 \right),
\eary \\
&& \label{tildeepsilon}
{\tilde \vep} =
\frac{- i\,m^{\prime}}{(M_1-M_2)-i(\Gamma_1-\Gamma_2) / 2}, \\
&& M_1-M_2 \approx M_S-M_L = -\Delta M_K, \nonumber \\
&& \Gamma_1-\Gamma_2 \approx \Gamma_S-\Gamma_L \approx
\Gamma_S. \nonumber
\eea
The observations give ${\tilde \vep} \simeq 2 \times 10^{-3}$
which then determines $m^{\prime}$ from which Eq.~(\ref{Gsw}) follows.

The superweak theory made three predictions:
\begin{enumerate}
\item The CP violation is completely described by ${\tilde \vep}$;
in particular the observables
\bean
\eta_{+-}=\eta_{00}={\tilde \vep},
\eean
where
\bean
\eta_{+-} &=& A(K_L \to \pi^+\,\pi^-)/A(K_S \to \pi^+\,\pi^-), \\
\eta_{00} &=& A(K_L \to \pi^0\,\pi^0)/A(K_S \to \pi^0\,\pi^0).
\eean
\item The phases of $\eta_{+-}$ and $\eta_{00}$ are equal and
determined by Eq.~(\ref{tildeepsilon}) to be about $43.5^{\circ}$
using the empirical values of $\Delta M_K$ and $\Gamma_S$.
\item The $CP$ violation is so small that it will not be seen anywhere
else.
\end{enumerate}
These predictions proved all too true for more than 25 years.

The alternative to superweak says that $CP$ is violated in the decay
amplitude $A_0\,e^{i\,\delta_0}$ and $A_2\,e^{i\,\delta_2}$
corresponding to final $I=0$ and $I=2$ $\pi\,\pi$ states, where
$\delta_I$ is the strong phase shift.  From $CPT$ and unitarity ${\rm
Im}\,A_I$ give the amplitudes for the $CP$-violating transitions $K_2
\to \pi\,\pi$.  Since there is still in any model the contribution
$m^{\prime}$ there are now three $CP$-violating quantities \cite{BS1965}
\bean
m^{\prime},\;\im\,A_0,\;\im\,A_2.
\eean
However there is a choice of phase convention using the $U(1)$
transformation $s \to s\,e^{-i\,\alpha}$ under which the strong and
electromagnetic interactions are invariant.  For the infinitesimal
$U(1)$ as an example
\bean
s &\to& s(1-i\,\alpha), \\
\im\,A_I &\to& \im\,A_I-\alpha\,\re\,A_I, \\
m^{\prime} &\to& m^{\prime}+\alpha(M_1-M_2).
\eean
Thus there are only two independent quantities which may be chosen as
\be
\bary{rcl}
\label{epsilons}
\vep^{\prime} &\propto& \im\,A_0/\re\,A_0 - \im\,A_2/\re\,A_2, \\
\vep &\simeq& {\tilde \vep} +i(\im\,A_0/\re\,A_0).
\eary
\ee
Then
\be
\bary{rcl}
\eta_{+-} &=& \vep +
\displaystyle{\frac{\vep^{\prime}}{1+\omega/\sqrt{2}}}, \\
\eta_{00} &=& \vep -
\displaystyle{\frac{2\,\vep^{\prime}}{1+\sqrt{2}\,\omega}},
\eary
\ee
where $\omega=\re\,A_2/\re\,A_0 \simeq 0.045$ and the last numerical
result is empirical and a sign of the $\Delta I=1/2$ rule.  The
quantity $\vep^{\prime}$ is a measure of $CP$ violation in the
decay amplitudes and so not due to a superweak interaction.

\section{The Standard Model vs Superweak}

For 35 years experiments have sought to determine $\vep^{\prime}$
by finding the difference between $\eta_{+-}$ and $\eta_{00}$ and thus
detecting $CP$ violation in the decay amplitude.  The experiments
actually measure
\be
\displaystyle{\left| \frac{\eta_{+-}}{\eta_{00}} \right|^2} \simeq 1+
6\,\re(\vep^{\prime}/\vep). \nonumber
\ee
By chance the phase of $\vep$ given to a good approximation by
Eq.~(\ref{tildeepsilon}) and the phase of $\vep^{\prime}$ given
by $(\pi/2-\delta_0+\delta_2)$ are both approximately equal to
$45^{\circ}$ so that $\re(\vep^{\prime}/\vep) \simeq
|\vep^{\prime}/\vep|$.  Recent experiments have left no doubt
that $|\vep^{\prime}/\vep|$ is non-zero and my average of the
results is
\bean
|\vep^{\prime}/\vep| = (20 \pm 6)\times 10^{-3}.
\eean

The big development in weak interactions was the spontaneously broken
gauge theory.  Originally given by Weinberg as a theory of leptons it
was extended by Glashow, Iliopoulos and Maiani (GIM) to quarks by
adding a fourth quark charm.  With the discovery of neutral currents
in 1973 this became the new Standard Model.  A striking feature of
this new theory was that it had $CP$ and $T$ invariance and so
provided no solution to the $CP$ violation problem.  In one paragraph
of a paper \cite{KM1973} in 1973 that few people noticed for several
years it was proposed that if there were six quarks then $CP$
violation could be allowed in the Standard Model.  With the discovery
of $b$ quark physics in 1978 this became the standard CKM model of
$CP$ violation.

The only place one can insert complex phases to give $CP$ violation in
the Standard Model is in the Yukawa interaction.  This shows up after
symmetry breaking in the unitary CKM matrix $V_{ji}$ connecting down
quarks $d_i$ to up quarks $u_j$ in the interaction with the $W$
bosons:
\be
{\bar u_j}\,V_{ji}\,\gamma^{\mu}(1-\gamma_5)d_i\,W_{\mu}. \nonumber
\ee
With only four quarks the unitary $2\times2$ matrix has three phases
which can be removed by the $U(1)$ transformations of $S$, charm $C$,
and electric charge $Q$.  With six quarks the $3\times3$ matrix has
six phases but only five can be removed.  In a standard phase
convention there are two matrix elements with large phases:
\be
\bary{rcl}
V_{ub} &=& |V_{ub}|e^{-i\,\gamma} \simeq A\,\lambda^3(\rho-i\,\eta), \\
V_{td} &=& |V_{td}|e^{-i\,\beta} \simeq A\,\lambda^3(1-\rho-i\,\eta),
\eary
\ee
where the last equality corresponds to a standard parameterization in
powers of $\lambda = \sin \theta_c \simeq 0.22$ (see the lectures by
Falk).

The $CP$ violation is directly proportional to $\eta$.  The
$CP$-violating observable $\vep$ (determined from $\eta_{+-}$) is
calculated from the box diagram giving the $K-{\bar K}$ mixing
parameter $m^{\prime}$.  (Note $\im\,A_0/\re\,A_0$ is very small
compared to ${\tilde \vep}$ in using Eq.~(\ref{epsilons}).)  This
is directly proportional to $\eta$ and requires that $\eta$ be between
$0.2$ and $0.6$.  This is very consistent with the constraint from the
magnitude of $|V_{ub}|$.  Note this consistency is significant; when
one thought $m_t$ was $20\,{\rm GeV}$ instead of $170\,{\rm GeV}$
the value of $\eta$ required was larger and the consistency was much
less obvious.  Given this value of $\eta$ the phase $\gamma$ is
expected to lie between $45^{\circ}$ and $135^{\circ}$ and the phase
$\beta$ between $15^{\circ}$ and $30^{\circ}$.

The calculation of $\vep^{\prime}$ in the Standard Model is very
uncertain due to the difficulty of calculating hadronic matrix
elements (see the lectures by Buchalla).  The dominant contribution is
the so-called penguin diagram in which a loop involving $t$ or $c$
quarks emits a gluon.  The $s \to d + g$ transition contributes only
to $\im\,A_0$.  There is also an electroweak penguin $s \to d +
(\gamma\;{\rm or}\;Z)$ which can contribute to $\im\,A_2$ and somewhat
decreases $\vep^{\prime}$ (see Eq.~(\ref{epsilons})).  Different
theoretical calculations give values of $\vep^{\prime}/\vep$
from $1.3\;{\rm to}\;6 \times 10^{-3} \eta$ with large errors.  Given
the uncertainties it seems the Standard Model can explain, though not
really predict, the value of $\vep^{\prime}/\vep$.

The direct $CP$ violation indicated by $\vep^{\prime}/\vep$ is
a major blow to the superweak theory.  Before completely abandoning
superweak I thought it would be good to define it.  There are many
possible theories which are effectively superweak; their common
feature is that at a low energy scale their effective interaction is
given by
\bean
{\cal H}_{eff} = {\cal H}_0 + {\cal H}_{sw},
\eean
where ${\cal H}_0$ is the Standard Model with the $CP$-violating
parameter $\eta$ chosen to zero so that all $CP$ violation is given by
${\cal H}_{sw}$ which contains terms of the form
\bean
G_{ijkl}\,{\bar q_i}\,q_j\,{\bar q_k}\,q_l
\eean
with all $G_{ijkl} \ll G_F$.  In particular, as noted, to explain
$\vep$, $G_{{\bar s}d{\bar s}d} \sim 10^{-10}\;{\rm
to}\;10^{-11}\,G_F$.  In general, the terms in ${\cal H}_{sw}$ are
derived from diagrams involving heavy particles beyond those in the
Standard Model \cite{B1986}.

Because $\vep^{\prime}$ is only about $5 \times 10^{-6}$ the
possibility arises that in some theories that are effectively
superweak such a value of $\vep^{\prime}$ could be explained.  In
fact as long as one is only dealing with the first four quarks one can
think of the Standard Model as effectively superweak.  Then ${\cal
H}_0$ is the $4$-quark Standard Model, which, as we noted, is
$CP$-invariant and ${\cal H}_{sw}$ corresponds to effective
$CP$-violating interactions due to diagrams involving the heavy $t$
quark which determines $\vep$ and $\vep^{\prime}$.

The really distinctive features of the CKM model appear only when we
consider the $CP$ violation involving $B$ mesons.  The $B-{\bar B}$
mixing is dominated by the box diagram involving $t$ quarks giving in
the $B-{\bar B}$ representation
\bean
M_{12} \propto (V_{td}^*\,V_{tb})^2 \simeq A^2\,\lambda^6
[(1-\rho)^2+\eta^2]e^{2i\,\beta}.
\eean
Thus in this phase convention there is large $CP$ violation in the
mixing.  The time-dependent $B^0({\bar B^0})$ decay rate to a $CP$
eigenstate has a term \cite{Q1998}
\be
\pm \eta_f\,\sin 2(\beta-\varphi_f)\sin \Delta m\,t,
\ee
where $\varphi_f$ is the $CP$-violating phase in the decay amplitude,
$\eta_f$ is the $CP$ eigenvalue, and $+(-)$ corresponds to initial
$B^0({\bar B^0})$.  The observation of this asymmetry is the major
goal of $B$ factories.  The first example will be the decay $B^0({\bar
B^0}) \to \Psi\,K_S$ which is due to the transition $b \to c\,{\bar
c}\,s$, for which $\varphi_f=0$.  A result from the hadron collider
experiment CDF has given the result $\sin 2\beta=0.8 \pm 0.4$.  By the
time these lectures are published more accurate values should be
available from the BABAR and BELLE experiments.

Large positive value of $\sin 2\beta$ will give strong qualitative
evidence for the Standard Model.  Nevertheless this asymmetry can
still be blamed on superweak mixing with $G_{{\bar b}d{\bar b}d} \sim
10^{-7}$, although only in special models would one expect such a
large effect in $B-{\bar B}$ mixing.  To finally kill all
superweaklike theories one must look for the expected large $CP$
violation in the $B$ decay amplitude.  To allow for the superweak
possibility we call the result of the first experiment $\sin 2{\tilde
\beta}$; then if a different decay to a final state $f^{\prime}$ gives
the asymmetry $\sin 2({\tilde \beta}-\varphi_{f^{\prime}})$, the phase
$\varphi_{f^{\prime}}$ is the relative phase of $B \to f^{\prime}$ to
that of $B \to \Psi\,K_S$.

The most obvious choice for $f^{\prime}$ is the decay $B \to
\pi^+\,\pi^-$.  In the ``tree'' approximation this is due to $b \to
u\,{\bar u}\,d$ and so that the phase
$\varphi_{f^{\prime}}=\varphi_{\pi}=-\gamma$.  For example, if
$\gamma=90^{\circ}$ then $\sin 2({\tilde \beta}+\gamma)=-\sin 2{\tilde
\beta}$, a very large direct $CP$-violating effect.  This I call the
$\vep^{\prime}$ experiment for the $B$ system.  In contrast to
$\vep^{\prime}$ for the $K$ system which is $5 \times 10^{-6}$ we
expect here $\sin 2({\tilde \beta}-\varphi_{\pi}) - \sin 2{\tilde
\beta}$ of order unity!

However, this may not prove so easy.  The branching ratio $B^0 \to
\pi^+\,\pi^-$ may be only about $5 \times 10^{-6}$.  Furthermore if
$\varphi_{\pi} = 2 {\tilde \beta}-\pi/2$ then $\sin 2({\tilde
\beta}-\varphi_{\pi}) = \sin 2{\tilde \beta}$ by accident.  In the
tree approximation if $2 {\tilde \beta} = 45^{\circ}$ this would occur
for $\gamma=45^{\circ}$ near the end of the allowed region.  In fact,
however, one expects a large contribution from ``penguin'' diagrams
corresponding to $b \to d + gluon$ via a $t$ quark loop (see the
lectures of Rosner).  Thus, the decay amplitude is given as
\be
\label{ampgeneral}
A(B \to \pi^+\,\pi^-) = T\,e^{-i\,\gamma} + P\,e^{i\,\beta}\,e^{i\,\Delta},
\ee
since the penguin is proportional to $V_{tb}\,V_{td}^*$ and so has the
phase $\beta$.  Estimates based on the rate of $B \to K\,\pi$
\cite{SW1994} suggest $P/T$ as high as $0.4$.  Assuming the strong
phase $\Delta$ is small, then $|\varphi_{\pi}|$ is less than
$|\gamma|$ so that $\varphi_{f^{\prime}}=-45^{\circ}$ corresponds to
$\gamma \sim 65^{\circ}$.  For values of $\gamma < 75^{\circ}$ it may
be difficult to detect the difference between $\sin 2{\tilde \beta}$
and $\sin 2({\tilde \beta}-\varphi_{\pi})$; since present evidence
from the ratio of $\Delta m(B_s)/\Delta m(B_d)$ suggests $\gamma <
90^{\circ}$, the region $\gamma < 75^{\circ}$ corresponds to most of
the range of $\gamma$.  It may prove easier then to consider other
decays such as $B^0 \to \rho^+\,\pi^-$ and $\rho^-\,\pi^+$.

An alternative would be to consider decays that are dominated by the
$b \to d$ penguin.  In this case $\varphi_f=\beta$ and $\sin 2({\tilde
\beta}-\varphi_f) = \sin 2({\tilde \beta}-\beta)$.  In the Standard
Model ${\tilde \beta} = \beta$ and so there is no asymmetry.  The
observation of such a zero asymmetry in contrast to the large
asymmetry for $B^0 \to \Psi\,K_S$ would then be evidence for a large
direct $CP$ violation.  Unfortunately the best candidates for such
decays might not be very practical; they correspond to $b \to d\,{\bar
s}\,s$ yielding $B^0 \to K_S\,K_S$, $B^0 \to \rho^0\,\eta$, etc.  Note
that even if the decay were not pure penguin one would expect an
asymmetry very different from that for $B^0 \to \Psi\,K_S$.

Another possibility is a $CP$-violating effect in which mixing is not
involved.  One can look for a difference between the rates of $B^+ \to
f$ and $B^- \to {\bar f}$ or in the case of $B^0$ looking at time
$t=0$ before mixing for the difference between $B^0 \to
f$ and ${\bar B^0} \to {\bar f}$.  In practice this means looking for
the $\cos \Delta m_B\,t$ term \cite{Q1998} rather than the $\sin
\Delta m_B\,t$.  As an example, consider $B^0 ({\bar B^0}) \to
\pi^+\,\pi^-$; from Eq.~(\ref{ampgeneral})
\bea
A_{\pi} &=&
\frac{\Gamma(B^0 \to \pi^+\,\pi^-)-\Gamma({\bar B^0} \to
\pi^+\,\pi^-)}
     {\Gamma(B^0 \to \pi^+\,\pi^-)+\Gamma({\bar B^0} \to
\pi^+\,\pi^-)}, \nonumber \\
\label{Api}
 &=&
-\frac{2\,T\,P\,\sin(\beta+\gamma)\sin\Delta}{T^2+P^2+2\,T\,P\,\cos(\beta+\gamma)\cos\Delta}.
\eea
For $\beta+\gamma \simeq 90^{\circ}$ this gives (with $P/T \equiv r$)
\bean
A_{\pi} = -\frac{2\,r}{1+r^2} \sin\Delta.
\eean
Thus a very large $CP$-violating asymmetry is possible if $r$ is
greater than $0.3$, but it all depends on the strong phase $\Delta$.

It is very difficult to make definite statements about the strong
phases in $B$ decays in contrast to $K$ decays.  For $K \to \pi\,\pi$
the final state interaction can be thought of as elastic scattering
with phase shifts $\delta_2$ and $\delta_0$ corresponding to
$\pi\,\pi$ states with $I=2$ and $I=0$.  However, $\pi\,\pi$ $s$-wave
scattering at $5$ GeV is highly inelastic involving many channels.
The phase $\Delta$ arises from the absorptive parts of diagrams
corresponding to the strong scattering from other final states into
the $\pi\,\pi$ state.  For any weak interaction operator ${\cal O}_i$
we can define the real decay amplitude in lowest order
\bean
M_{fi} = M_{fi}^0 = \lan f | {\cal O}_i | B \ran.
\eean
If $f$ were an eigenstate one would then multiply this by
$e^{i\,\delta_f}$.  Going from the eigenstate basis to the states of
interest
\be
\label{Mfi}
M_{fi} = \sum_{f'} \lan f | S^{1/2} | f' \ran \lan f' | {\cal O}_i | B
\ran,
\ee
where $S$ is the strong-interaction $S$ matrix.

The sum in Eq.~(\ref{Mfi}) is over a large number (almost uncountable)
of states.  One can only make some general comments about it:
\begin{enumerate}

\item The strong phase depends on the operator ${\cal O}_i$ that
affects the relative importance of different states $f'$.  The phase
$\Delta$ in Eqs.~(\ref{ampgeneral}) and (\ref{Api}) is the difference
between the strong phase of the ``tree'' operator and that of the
``penguin''.

\item Since the strong scattering is expected to be very inelastic the
diagonal element $\lan f | S^{1/2} | f \ran$ has as its major effect
the reduction of $M_{fi}$; this is a kind of absorption effect.  Thus
we could write
\bean
M_{fi} = M_{fi}^0 a_i + i\,R_i = |M_{fi}| e^{i\,\delta_i},
\eean
where $a<1$ is the reduction due to absorption.  For a ``typical''
state, by unitarity, the scattering ``in'' due to $R$ compensates for
the scattering ``out'' so
\be
R_i = \sqrt{1-a_i^2} = \sin \delta_i.
\ee

\item An estimate can be based on a crude statistical argument
\cite{SW1999} in which case one can reduce the multichannel problem to
an equivalent 2-channel problem
\bean
S =
\left(
\bary{cc}
\cos2\theta & i\,\sin2\theta \\
i\,\sin2\theta & \cos2\theta
\eary
\right).
\eean
For $f=\pi\,\pi$ the diagonal element $\cos2\theta$ can be estimated
by extrapolating data from $\pi\,N$ scattering to $\pi\,\pi$ giving
\bean
\cos2\theta = \eta \simeq 0.7.
\eean
Then Eq.~(\ref{Mfi}) becomes
\bea
M_{1i} &=& \cos\theta\,A_{1i} + i\,\sin\theta\,A_{2i},
\nonumber \\
\label{tandelta}
\tan \delta_i &=& i\,\tan\theta\,\frac{A_{2i}}{A_{1i}},
\eea
where $\tan\theta = \left( \frac{1-\cos2\theta}{1+\cos2\theta}
\right)^{1/2} \simeq 0.42$.

The ``typical'' result corresponds to $A_2 = A_1$ and gives a strong
phase of $20^{\circ}$ to $25^{\circ}$.  The quantitative conclusion
from Eq.~(\ref{tandelta}) is that if the state of interest (labeled
$1$ here) is a ``more probable'' final state than the states into
which that state scatters (lumped into state 2 here) then the strong
phase may be small.

\end{enumerate}

In conclusion, after it is clearly shown that $CP$ violation is not
superweak the next step is to find quantitative tests of the Standard
Model by showing the consistency of a number of different
experiments.  This is a major program for the next decade.

\section{Time Reversal Violation}

By the $CPT$ theorem $CP$ violation implies time-reversal violation
(TRV).  Strong evidence for $CPT$ invariance comes from the phase of
$\vep$ determined from Eq.~(\ref{tildeepsilon}).  $CPT$ violation
would allow a real off-diagonal term $m^{\prime\prime}$ in the matrix
in addition to $i\,m^{\prime}$ and thus would change the phase.  Since
the measured phase agrees with theory to about $1^{\circ}$ there is a
strong limit on $m^{\prime\prime}$ which corresponds to a limit on
$[m(K^0)-m({\bar K^0})]/m_K \leq 10^{-18}$.  Nevertheless, it is of
great interest to look for direct evidence for TRV both as another way
to study $CP$ violation as well as a way to demonstrate $T$ violation
in a straightforward way \cite{W1999}.

We discuss here four types of direct evidence for TRV; by this I mean
a single experiment that by itself is seen to violate $T$.  These are
\begin{enumerate}

\item A non-zero value of a $T$-odd observable in a stationary state.
The simplest example is the electric dipole moment of an elementary
particle or an atom.

\item A violation of the reciprocity condition on the $S$ matrix
\bean
S_{fi} = S_{-i,-f}
\eean
corresponding to comparing a reaction and its inverse.

\item A non-zero value of a $T$-odd observable in the final state of a
weak decay.  As discussed below this depends upon the neglect of
final-state interactions.

\item In an oscillation a difference in the probability of $a \to b$
from $b \to a$ at a given time.  It is interesting to note that each
example immediately implies a test of $CP$ violation (conceptual if
not practical) by going to the anti-particles.  In contrast the
simplest tests of $CP$ violation have no direct relation to TRV; for
example, $\Gamma(B \to f) \!= \Gamma({\bar B} \to {\bar f})$ involves
a rate which has nothing to do with a TRV observable.

\end{enumerate}

Experimental limits on the dipole moments of the electron and neutron
are
\bean
d_n &\leq& 10^{-25}\; {\mathrm e-cm}, \\
d_e &<& 4 \times 10^{-27}\; {\mathrm e-cm}.
\eean
In the Standard Model $d_n$ is second-order weak and the calculation
depends on long-distance effects giving of order $10^{-32}$ e-cm;
$d_e$ is third order and perhaps $10^{-38}$ e-cm.  Thus the interest
lies in the search for physics beyond the Standard Model (see the
lecture by Thomas).

Many tests of the reciprocity relations exist for strong interactions
although they are of limited accuracy; it is very hard to study the
reverse of weak interactions.  An interesting proposal by Bowman
involves the scattering of slow neutrons from polarized nuclei in the
resonance region.  One compares the observable $< {\vec \sigma_n} \cdot
{\vec I} \times {\vec k} >$ for incident polarization ${\vec \sigma_n}$
and final polarization ${\vec \sigma_n}$, where ${\vec I}$ is the
nuclear polarization.  $T$-violating effects in the nuclear wave
functions could enhance the result.

An example of a ``$T$-odd observable'' in the final state of decay is
the muon polarization
\bean
P = < {\vec \sigma^{\mu}} \cdot {\vec k_{\mu}} \times {\vec k_{\nu}} >
\eean
in the decay $K \to \pi + \mu + \nu$.  This does not really violate
$T$ except in the Born approximation when final state interactions
(FSI) can be avoided.  As a simple didactic example, consider the
scattering from a potential
\bean
V_0 + V_1\,{\vec \sigma} \cdot {\vec L}
\eean
which certainly is $T$-invariant.  The resulting amplitude is
\bean
A = f_0 + i\,f_1\,{\vec \sigma} \cdot {\hat n}
\eean
where ${\hat n}$ is the normal to the scattering plane.  In the Born
approximation $f_0$ and $f_1$ are real and so $<{\vec \sigma} \cdot
{\hat n}> = 0$, but beyond the Born approximation $f_0$ and $f_1$ are
complex; for example, if $s$ and $p$ waves dominate $f_0$ would have a
phase $e^{i\,\delta_0}$ and $f_1$ the phase $e^{i\,f_1}$.

For the case of $K^0 \to \pi^+ + \mu^- + {\bar \nu_{\mu}}$ there is a
Coulomb FSI so that without $T$ violation, $P \sim 10^{-3}$; for the
case of $K^+ \to \pi^0 + \mu^+ + \nu_{\mu}$ the FSI involves $2\gamma$
exchange so that $P \sim 10^{-6}$.  In the Standard Model the real TRV
is expected to vanish in semi-leptonic decays.  One would expect a
real TRV in non-leptonic decays such as $\Lambda \to p\,\pi^-$ where
there is a defined parameter
\bean
\beta \propto <{\vec \sigma^{\Lambda}} \cdot {\vec \sigma^p} \times {\vec k}>.
\eean
However, the FSI effect is much larger being proportional to
$\sin(\delta_p-\delta_s)$ where $\delta_p$, $\delta_s$ are the $\pi$-$p$
phase shifts.  If the experiment is also done with ${\bar \Lambda}$,
then $\beta+{\bar \beta}$ is a clear $CP$-violating effect and is
associated with true TRV, but this is hardly ``direct evidence'' of
TRV.

A large ``$T$-odd observable'' has been found in the decay $K_L \to
\pi^+\,\pi^-\,e^+\,e^-$
\bean
C = < {\hat n_e} \times {\hat n_{\pi}} \cdot {\hat z} >\,
<{\hat n_e} \cdot {\hat n_{\pi}}>,
\eean
where ${\hat n_e} ({\hat n_{\pi}})$ are the normals to the
$e^+\,e^-(\pi^+\,\pi^-)$ planes and ${\hat z}$ is the unit vector
between the pairs.  This was predicted as a result of $K-{\bar K}$
mixing as an interference between an $M1$ virtual $\gamma$ from $K_2
\to \pi\,\pi\,\gamma$ and an $E_1$ virtual bremsstrahlung from $K_1
\to \pi\,\pi\,\gamma$.  The theoretical result \cite{SW1992} is
\bean
C = 0.15\,\sin(\varphi_{\vep}+\Delta),
\eean
where $\Delta \simeq 30^{\circ}$ comes from $\pi\,\pi$ phase shifts.
The experimental result \cite{A2000} verifies this; the result is so
large because for the $e^+\,e^-$ energy considered the $E1$ is much
larger than $M1$ which compensates for the small admixture
$|\vep|$ of $K_1$.  Since $\Delta$ is involved this is not again
obvious TRV.  It is of didactic interest to consider the limit $\Delta
\to 0$.  In this case $C$ is proportional to $\sin
\varphi_{\vep}$.  We know that $\varphi_{\vep} \simeq
\frac{\pi}{4}$ in accordance with $CPT$ invariance and $T$ violation.
If we had assumed $CPT$ violation and $T$ invariance it follows from
an analysis like that of Eq.~(\ref{tildeepsilon}), replacing
$i\,m^{\prime}$ by $m^{\prime\prime}$, that $\varphi_{\vep} \simeq
\frac{3\pi}{4}$ and so we get the same value of $C$ even though there
is no TRV and no FSI!

The explanation lies in the fact that we are here sensitive to higher
order weak effects which show up in $\Delta\Gamma$ in
Eq.~(\ref{tildeepsilon}).  I call this the ``initial state
interaction''.  This could be considered as a contribution to the
decay amplitude of the form
\bean
K^0 \to \pi^+\,\pi^- \to {\bar K^0} \to \pi^+\,\pi^-\,e^+\,e^-
\eean
with $\pi^+\,\pi^-$ on the mass shell thus giving an absorptive part.
Only if $\Delta=0$ and $\Delta\Gamma=0$ would a non-zero $C$ directly
show TRV.

The possibility of seeing $CP$ violation in neutrino oscillations from
the difference between $\nu_{\mu} \to \nu_e$ and ${\bar \nu_{\mu}} \to
{\bar \nu_e}$ has been discussed in many papers (see the talk by
Kayser).  The same formula gives the TRV difference between $\nu_{\mu}
\to \nu_e$ and $\nu_e \to \nu_{\mu}$.  Note that the time dependence
(or, equivalently, the distance dependence) of the difference is an
odd function of time.  The possibility of doing the TRV experiment
requires beams of both $\nu_{\mu}$ and $\nu_e$ as has been proposed
for ``neutrino factories'' based on a muon storage ring.

In the $CP$ LEAR experiment \cite{A1998} a difference has been
observed between the transitions $K^0 \to {\bar K^0}$ and ${\bar K^0}
\to K^0$.  Here the initial $K^0\,({\bar K^0})$ has been identified
by its associated production with a $K^+\,(K^-)$ and the final ${\bar
K^0}\, (K^0)$ by the charge of the lepton in the semi-leptonic decay.
The result in agreement with a simple calculation is
\bean
\frac{\Gamma(K^0 \to {\bar K^0}) - \Gamma({\bar K^0}
\to K^0)}
{\Gamma(K^0 \to {\bar K^0}) + \Gamma({\bar K^0}
\to K^0)}
=4\,\re\,\vep
\eean
independent of time.  This seems somewhat strange since we expected an
odd function of time.  One can also ask from unitarity if $K^0$ goes
to ${\bar K^0}$ more than ${\bar K^0}$ goes to $K^0$ what compensates
for this.  The answer is that the ${\bar K^0}$ decays to $\pi\,\pi$
more than $K^0$.  Thus, decay plays an essential role, rendering this
as a direct test of TRV somewhat questionable.

As we have noted the phase of $\vep$ is completely consistent with
$CPT$ invariance.  There is no reason to doubt $CPT$ invariance, which
appears very fundamental, and so we conclude that the observed $CP$
violation is associated with TRV.  Nevertheless, unambiguous direct
tests of TRV may prove very difficult.

\section*{Acknowledgments}

Many of the early papers on $CP$ violation are reprinted in
L. Wolfenstein, {\it CP Violation} (North Holland) (1989).  This work
was supported by the U.S. Department of Energy under Grant
No. DE-FG02-91ER40682.

\end{document}